\documentstyle[epsf,11pt,a4,palatino]{article}
\input psfig.tex
\def\lsim{\lower.5ex\hbox{$\; \buildrel < \over \sim \;$}}
\def\gsim{\lower.5ex\hbox{$\; \buildrel > \over \sim \;$}}
\font\au=cmr10
\font\ad=cmti10 at 9pt

\def\pmb#1{\setbox0=\hbox{$#1$}%
\kern-.025em\copy0\kern-\wd0
\kern.05em\copy0\kern-\wd0
\kern-.025em\raise.0433em\box0}
\def\lsim{\lower.5ex\hbox{$\; \buildrel < \over \sim \;$}}
\def\gsim{\lower.5ex\hbox{$\; \buildrel > \over \sim \;$}}

\begin{document}
\noindent
\hrule
\begin {center}
{\large\bf Behaviour of relativistic black hole accretion sufficiently close to the horizon}
\end{center}
\hrule
\vskip 0.25cm
\begin {center}
{\au {\sf Paramita Barai$^1$, Ipsita Chakraborty$^2$, Tapas K Das$^{3,4}$, Paul J. Wiita$^{5,6}$}\\ [0.2truecm]
{\ad $^1$ Department de physique, Universite Laval, Quebec City, Canada, {\tt paramita.barai.1@ulaval.ca}\\
$^2$ Adamas Institute of Technology, Kolkata 700126, India, {\tt ipsita14@gmail.com}\\
$^3$ Harish Chandra Research Institute, Allahabad 211019, India, {\tt tapas@hri.res.in}\\
$^4$ Theoretical Institute for Advanced Research in Astrophysics, Taiwan {\tt tapas@tiara.sinica.tw.edu}\\
$^5$ Institute for Advanced Study, Princeton University, USA {\tt wiita@sns.ias.edu}\\
$^6$ Department of Physics and Astronomy, Georgia State University, USA {\tt wiita@chara.gsu.edu}}} 
\end{center}
\vskip 0.25truecm
\begin{abstract}
\noindent
This work introduces a novel formalism to investigate the role of the spin 
of astrophysical black holes in determining the behaviour of matter 
falling onto such accretors. Equations describing the general 
relativistic hydrodynamic accretion flow in the Kerr metric are formulated, and
stationary solutions for such flow equations are provided. 
The accreting matter may become multi-transonic, allowing a stationary shock to form for 
certain initial boundary conditions. Such a shock determines the disc geometry 
and can drive strong outflows. The properties of matter extremely close to
the event horizon are studied as a function of the Kerr parameter, leading 
to the possibility of detecting a new spectral signature of black hole spin. 
\end{abstract}
\vskip 0.25truecm
\noindent
\section{Flow Structure}
\noindent
The energy momentum tensor ${\Im}^{{\mu}{\nu}}$ has been formulated (the general 
most form of ${\Im}^{{\mu}{\nu}}$ is available in Novikov \& Thorne 1973)
in the Boyer-Lindquist
co-ordinates, and its covariant derivative has been evaluated to obtain the 
general relativistic Euler equation and the equation of continuity, in the form of 
a set of spatio-temporal first order differential equations. Axi-symmetry 
has been adopted for the flow geometry as described in figure 1, where, 
following Abramowicz, Lanza \& Percival (1997), the
disk height $h(r)$ is calculated as:
\begin{equation}
h(r)=\sqrt{\frac{2}{\gamma + 1}} r^{2} \left[ \frac{(\gamma - 1)c^{2}_{c}}
{\{\gamma - (1+c^{2}_{s})\} \{ \lambda^{2}v_t^2-a^{2}(v_{t}-1) \}}\right] ^{\frac{1}{2}} 
\label{eq1}
\end{equation}
Two first integrals of motion for the system defined along streamlines,
the conserved specific flow energy, ${\cal E}$ and the baryonic load rate,
${\dot M}$, along with the corresponding entropy accretion rate, ${\dot {\Xi}}$, are 
calculated as:
\begin{equation}
{\cal E} =
\left[ \frac{(\gamma -1)}{\gamma -(1+c^{2}_{s})} \right]
\sqrt{\left(\frac{1}{1-u^{2}}\right)
\left[ \frac{Ar^{2}\Delta}{A^{2}-4\lambda arA +
\lambda^{2}r^{2}(4a^{2}-r^{2}\Delta)} \right] } \, ,~~
{\dot M}=4{\pi}{\Delta}^{\frac{1}{2}}H{\rho}\frac{u}{\sqrt{1-u^2}} \, ,
\label{eq2}
\end{equation}
\begin{equation}
{\dot \Xi}
 = \left( \frac{1}{\gamma} \right)^{\left( \frac{1}{\gamma-1} \right)}
4\pi \Delta^{\frac{1}{2}} c_{s}^{\left( \frac{2}{\gamma - 1}\right) } \frac{u}{\sqrt{1-u^2}}\left[\frac{(\gamma -1)}{\gamma -(1+c^{2}_
{s})}
\right] ^{\left( \frac{1}{\gamma -1} \right) } H(r)
\label{eq3}
\end{equation}
Differential solutions of the first integrals of motion provides the three 
velocity gradient (of the flow) as a first order autonomous dynamical system:
\begin{equation}
\frac{du}{dr}=
\frac{\displaystyle
\frac{2c_{s}^2}{\left(\gamma+1\right)}
  \left[ \frac{r-1}{\Delta} + \frac{2}{r} -
         \frac{v_{t}\sigma \chi}{4\psi}
  \right] -
  \frac{\chi}{2}}
{ \displaystyle{\frac{u}{\left(1-u^2\right)} -
  \frac{2c_{s}^2}{ \left(\gamma+1\right) \left(1-u^2\right) u }
   \left[ 1-\frac{u^2v_{t}\sigma}{2\psi} \right] }},
\label{eq4}
\end{equation}
In Eqns. (\ref{eq1}--\ref{eq4}), $r,u,c_s,\rho,\lambda,\gamma,a$ are 
the radial distance on the equatorial plane, three (flow) velocity,
sound speed, density, flow specific angular momentum, flow adiabatic index and the 
black hole spin (the Kerr parameter) respectively. $A$ and $\Delta$ are
functions of $r$ and $a$, $v_t$ is function of $u,r,\lambda,a,A,\Delta$,
and other terms (especially in (\ref{eq4})) are functions of 
above quantities and of various metric elements as well as their 
space derivatives. Further details about the above mentioned equations 
are available in Barai, Das \& Wiita (2004).
\section{Transonicity, Shock and Outflow Generation}
\noindent
The fixed point solution obtained from (\ref{eq4}) provides the number of critical 
points the flow can pass through, and thus provides the number of transonic flips. 

Region of multi-transonicity for both accretion (marked A) and wind (marked W)
in the parameter space are shown in figure 2. The labels I and O indicate regions 
with lone inner and outer critical points, respectively. The plot has been 
generated for $a=0.3$ and $\gamma=1.33$. 

Multi-transonic (actually, bi-modal, since the middle sonic point does not 
support a steady solution that passes through it) accretion flows may encounter a
stationary shock. 
The general relativistic shock conditions are formulated:
\begin{equation}
\left[\left[{\rho}u\Gamma_{u}\right]\right]=0,
\left[\left[{\Im}_{t\mu}{\eta}^{\mu}\right]\right]=
\left[\left[(p+\epsilon)v_t u\Gamma_{u} \right]\right]=0,
\left[\left[{\Im}_{\mu\nu}{\eta}^{\mu}{\eta}^{\nu}\right]\right]=
\left[\left[(p+\epsilon)u^2\Gamma_{u}^2+p \right]\right]=0,
\label{eq5}
\end{equation}
where $\Gamma_u$ is the Lorentz factor, and $\eta_\mu$ is the normal
to the hypersurface $\Sigma$ of discontinuity. The expression for the 
shock invariant ${\cal S}_h$ is computed:
\begin{equation}
{\cal S}_h=
c_s^{\frac{2\gamma+3}{\gamma-1}}
\left(\gamma-1-c_s^2\right)^{\frac{3\gamma+1}{2\left(1-\gamma\right)}}
u\left(1-u^2\right)^{-\frac{1}{2}}
\left[\lambda^2v_t^2-a^2\left(v_t-1\right)\right]^{-\frac{1}{2}}
\left[
\frac{u^2\left(\gamma-c_s^2\right)+c_s^2}{c_s^2\left(1-u^2\right)}
\right]
\label{eq6}
\end{equation}
The shock strength and the entropy enhancement ratio (at the shock) are 
defined as the ratios of Mach numbers $\left(M_-/M_+\right)$ and the
entropy accretion rate $\left({\Xi}_+/{\Xi}_-\right)$ respectively, 
where $-$ and $+$ indicate the pre- and post-shock values of any 
flow variable. 

Simultaneous solution of Eqns. (\ref{eq1}--\ref{eq6}) provides the phase 
portrait of the shocked multi-transonic accretion, as shown in 
figure 3. Transonic flow (marked by `A') through the outer sonic 
point encounters a stable stationary Rankine-Hugoniot shock transition 
marked by S1 (S2 being the other formal shock location, which is unstable 
and hence is not considered), produces post shock subsonic flow, which again 
becomes supersonic (segment marked by `a') after passing through the inner 
sonic point; and finally plunges through the event horizon. 
As a consequence of the shock formation, the post shock flow has higher 
temperature, density, pressure and residence time compared to its
pre-shock counterpart. This favours the generation of an optically thick
halo (yellow coloured elliptic patches in figure 4, where the disc structure 
has been obtained by solving (\ref{eq1}-\ref{eq6})) in the 
post shock region as a result of disc evaporation,
and initiates the production of thermally and centrifugally driven cosmic
outflows. 

Figure 5a shows the variation of shock location $r_{sh}$ as a function of 
black hole spin. Figure 5b shows that the stronger shocks 
(higher value of $\left(M_-/M_+\right)$) are produced closer to 
the black hole, and is a result of maximum entropy production 
(measured by $\left({\Xi}_+/{\Xi}_-\right)$) due to the shock. 
Pre- and post-shock ratios of temperature, pressure, density 
and velocity (some of the normalized by appropriate numerical 
factors so as to fit in the same figure) as a function of black hole spin are plotted in 
figure 5c.
\section{Quasi-terminal Values}
Flow variables (calculated along the solution `a') at a very close 
proximity $r_\delta={r_+}+\delta$ ($r_+=1+\sqrt{(1-a^2)}$, and 
$\delta=0.001 r_g$) of the event horizon are termed as 
quasi-terminal values, and are distinguished with a subscript $\delta$.
The variation of the (normalized) quasi-terminal temperature (${\rm T}_\delta$),
density ($\rho_\delta$), pressure (${\rm p}_\delta$), as a function 
of the black hole spin, are shown in the figure 6. 
The calculations are performed for a $3{\times}10^6 M_{\odot}$ 
black hole accreting at a rate of $4.29{\times}10^{-6}
M_{\odot}{\mathrm Yr}^{-1}$.\\ \\
\noindent
Shock formation is naturally also possible
for other values of the black hole mass and the accretion rate, as
well as for values of the black hole spin higher than those
shown here in Fig. 6. Accretion onto the rotating black hole
with the Kerr parameter as high as $0.95$ has been observed to
form shocks; the details of such work will be reported elsewhere
(Barai, Chakraborty, Das \& Wiita, in preparation).
\section{Conclusion}
\begin{itemize}
\item For the first time, the shocked relativistic accretion in 
the Kerr metric has been studied at such a close proximity to the 
black hole event horizon. 
\item The effects of the black hole spin on the dynamical and
thermodynamical properties of the accreting material have been made
explicit.
\item Accurate estimates of the flow temperature, velocity and 
density profile has been made. This leads to a 
possible formalism for the study of the spectral signature of 
the black hole spin (work 
in progress).
\end{itemize}
\section*{Acknowledgements}
The work of TKD was partially supported by  the Theoretical
Institute for Advanced Research in Astrophysics (TIARA)
operated under Academia Sinica and the National Science
Council Excellence Projects program in Taiwan, administered
through grant NSC 96-2752-M-007-007-PAE. IC would like to
acknowledge the kind hospitality provided by HRI, Allahabad, India.

{}
\newpage
\begin{figure}
\vbox{
\vskip -3.0cm
\centerline{
\psfig{file=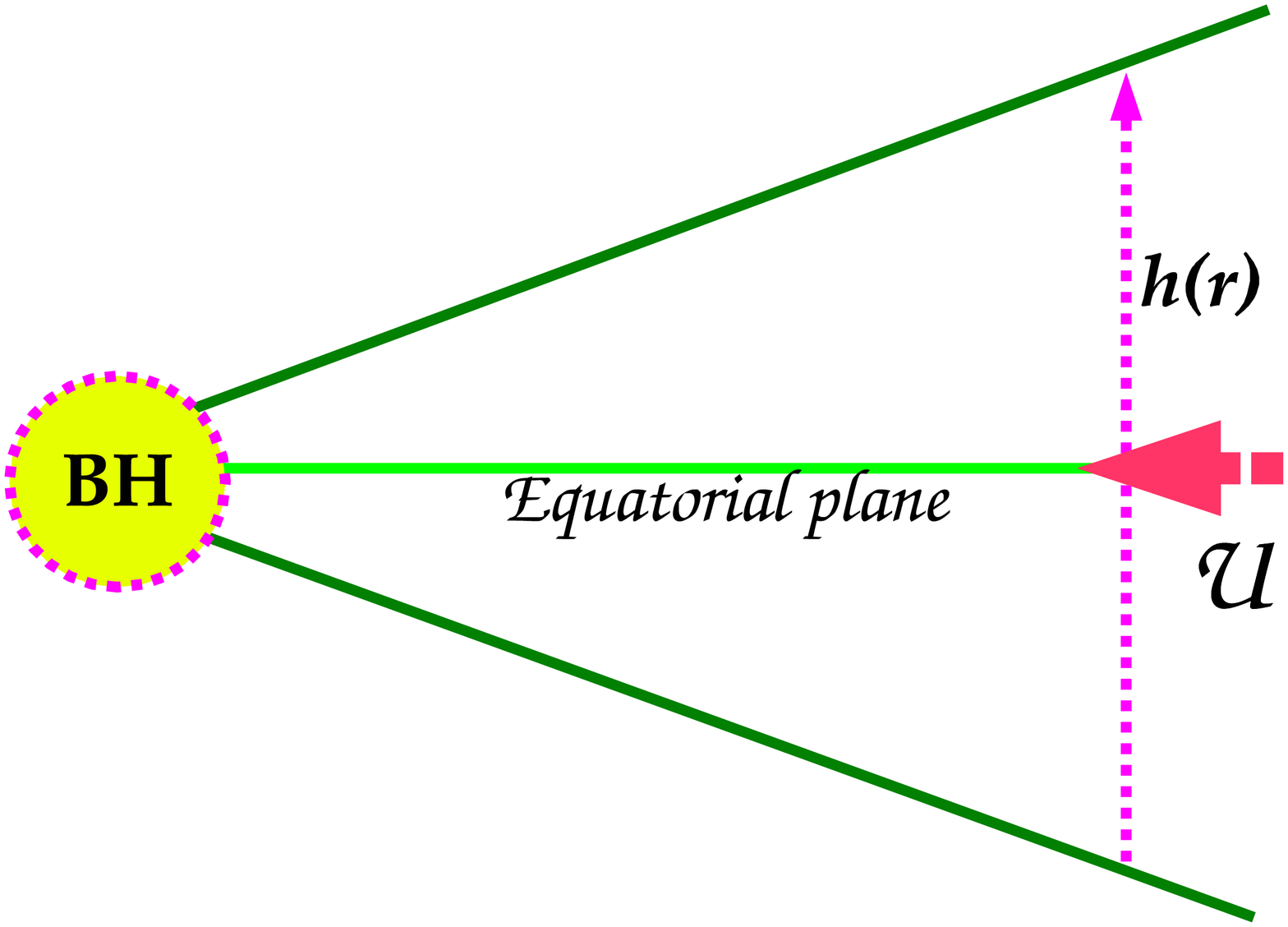,height=11cm,width=14.2cm,angle=0.0}}
{\bf Figure 1:}
The yellow circular patch with BH written inside represents the black hole
and the pink dashed boundary mimics the event horizon. The wedge shaped dark green
lines represents the envelop of the accretion disc. The light green line centrally
flanked by the two dark green disk boundaries, is the equatorial plane, on which all
of the dynamical quantities (e.g., the advective three velocity $u$) are assumed to be
confined. Any thermodynamic quantity (e.g., the flow density) is
averaged over the local disc height ${\bf h}(r)$.}
\end{figure}
\begin{figure}
\vbox{
\vskip -2.0cm
\centerline{
\psfig{file=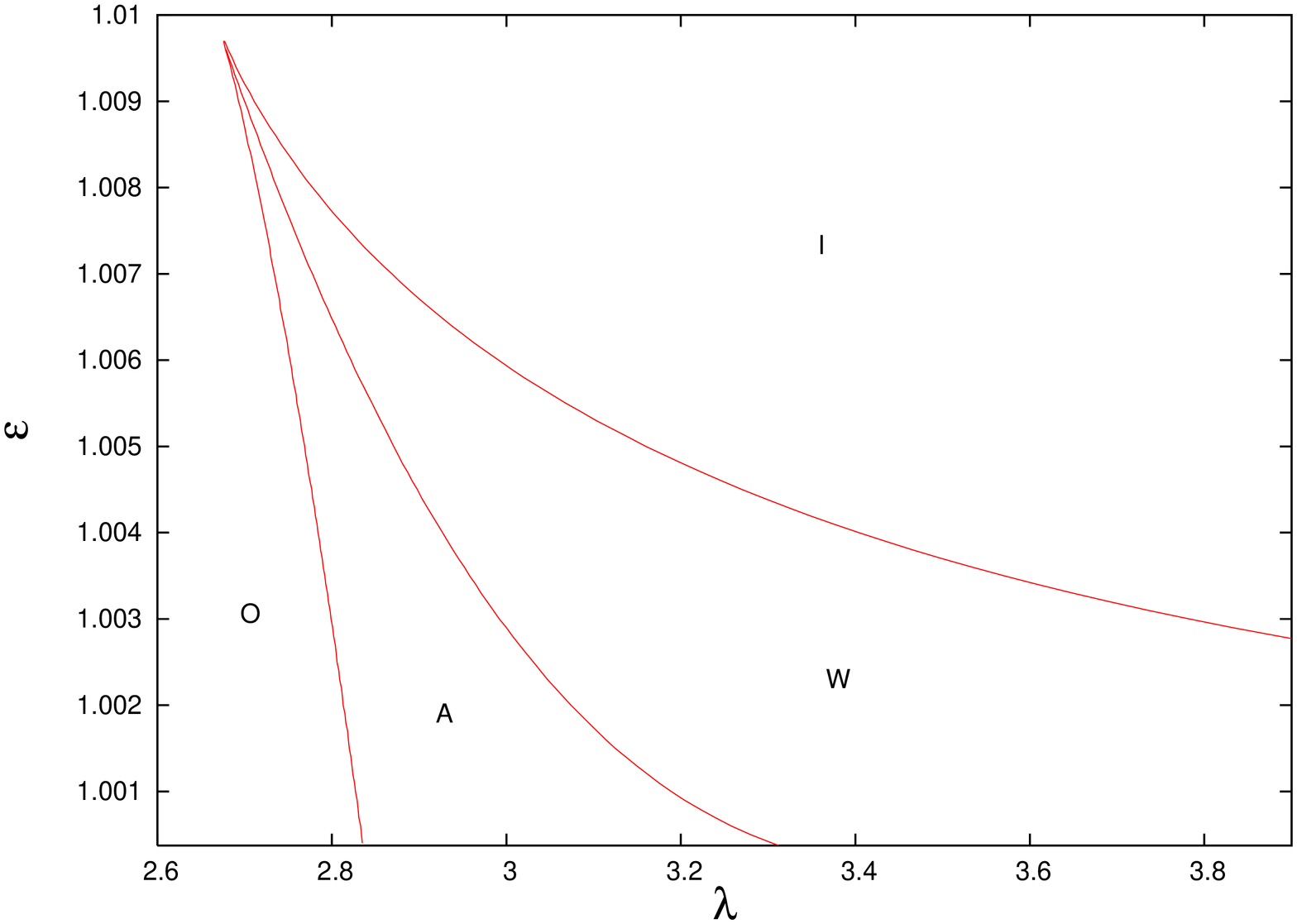,height=11cm,width=14.2cm,angle=270.0}}
{\bf Figure 2}
}
\end{figure}
\begin{figure}
\vbox{
\vskip -0.5cm
\centerline{
\psfig{file=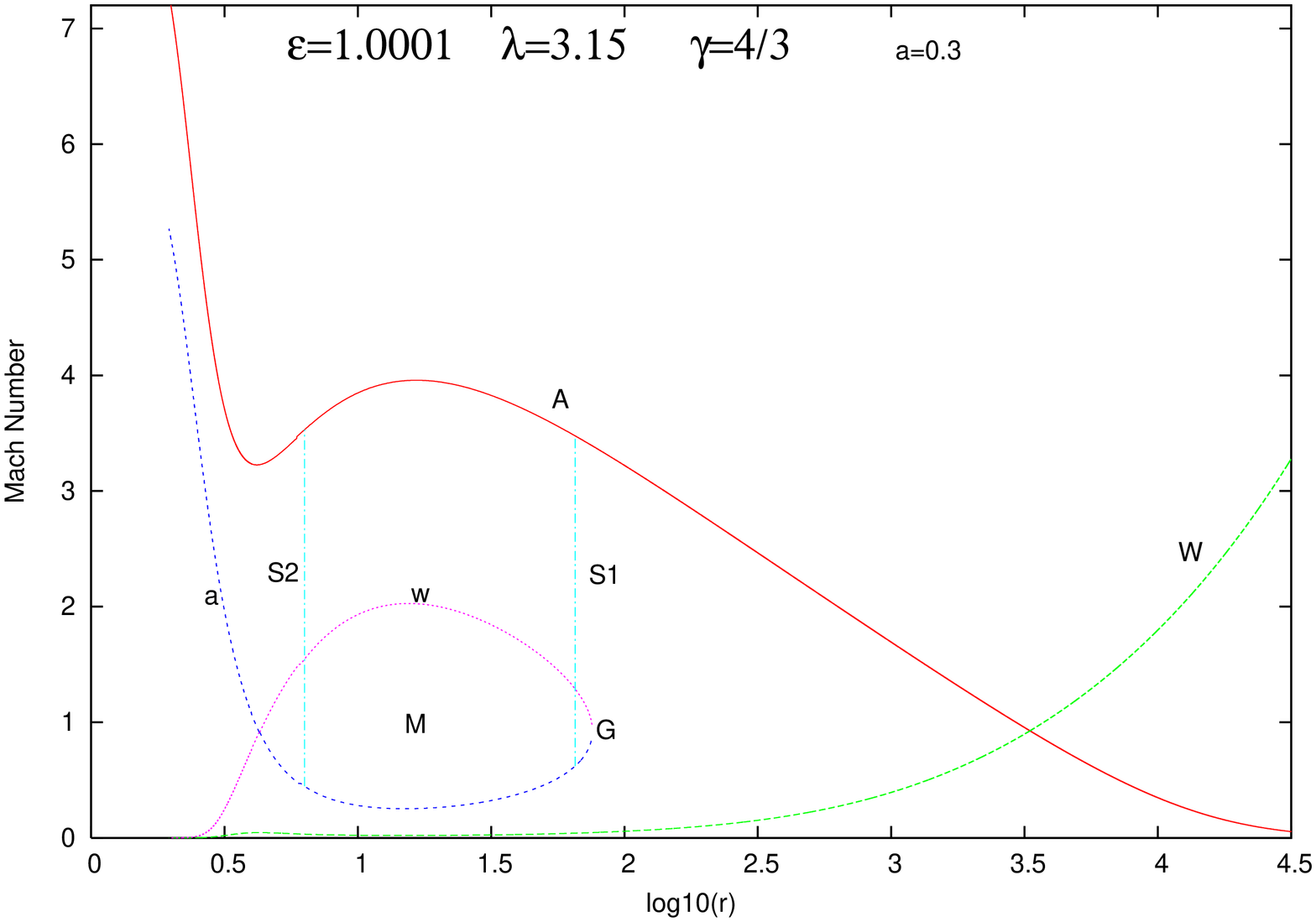,height=12cm,width=16.2cm,angle=270.0}}
{\bf Figure 3}
}
\end{figure}
\begin{figure}
\vbox{
\vskip -5.0cm
\centerline{
\psfig{file=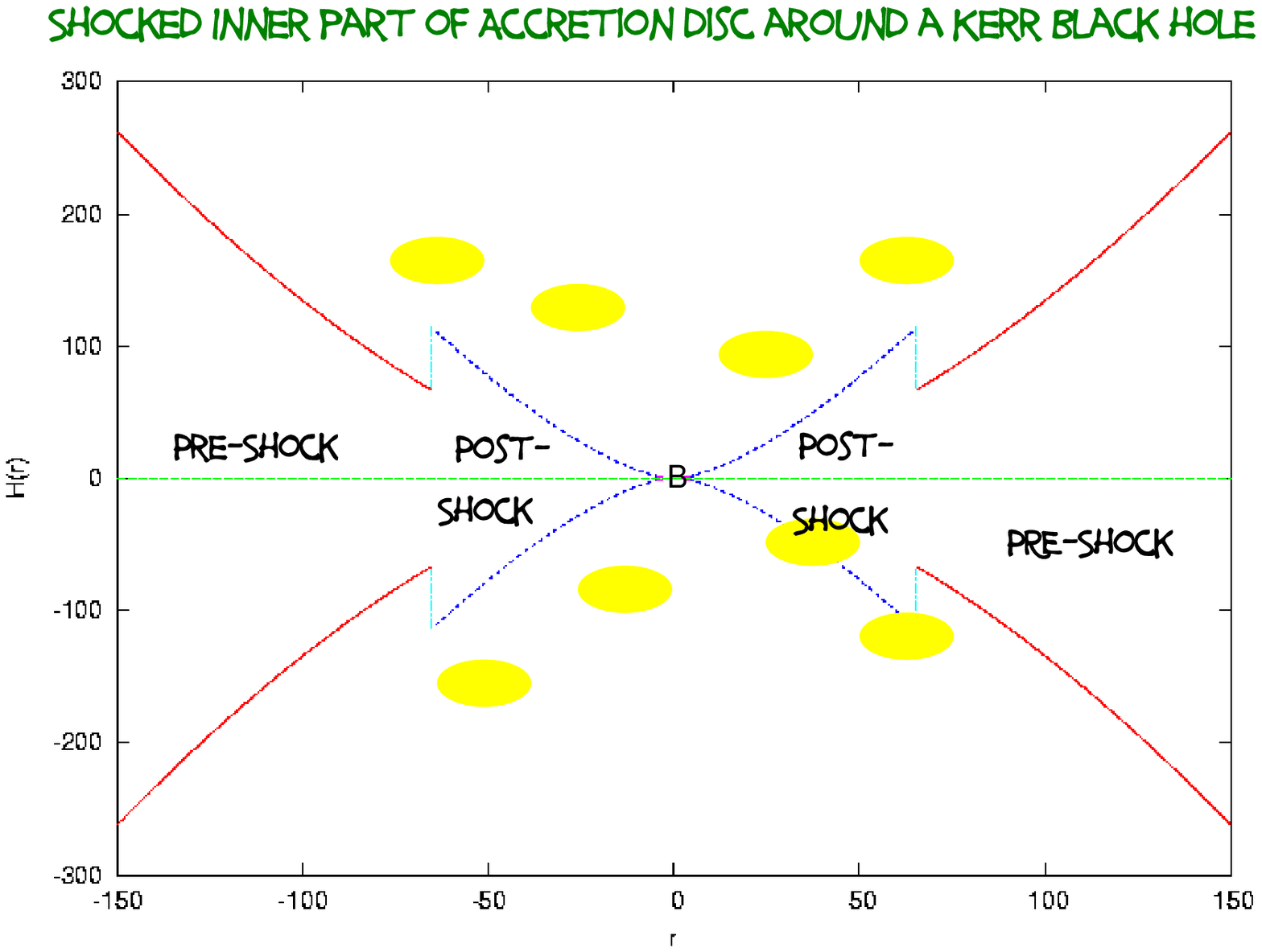,height=20cm,width=17.2cm,angle=0.0}}
\vskip -5.0cm
{\bf Figure 4:} Pre- and post-shock disc geometry with thermally driven 
optically thick halo.
}
\end{figure}
\begin{figure}
\vbox{
\vskip -9.0cm
\centerline{
\psfig{file=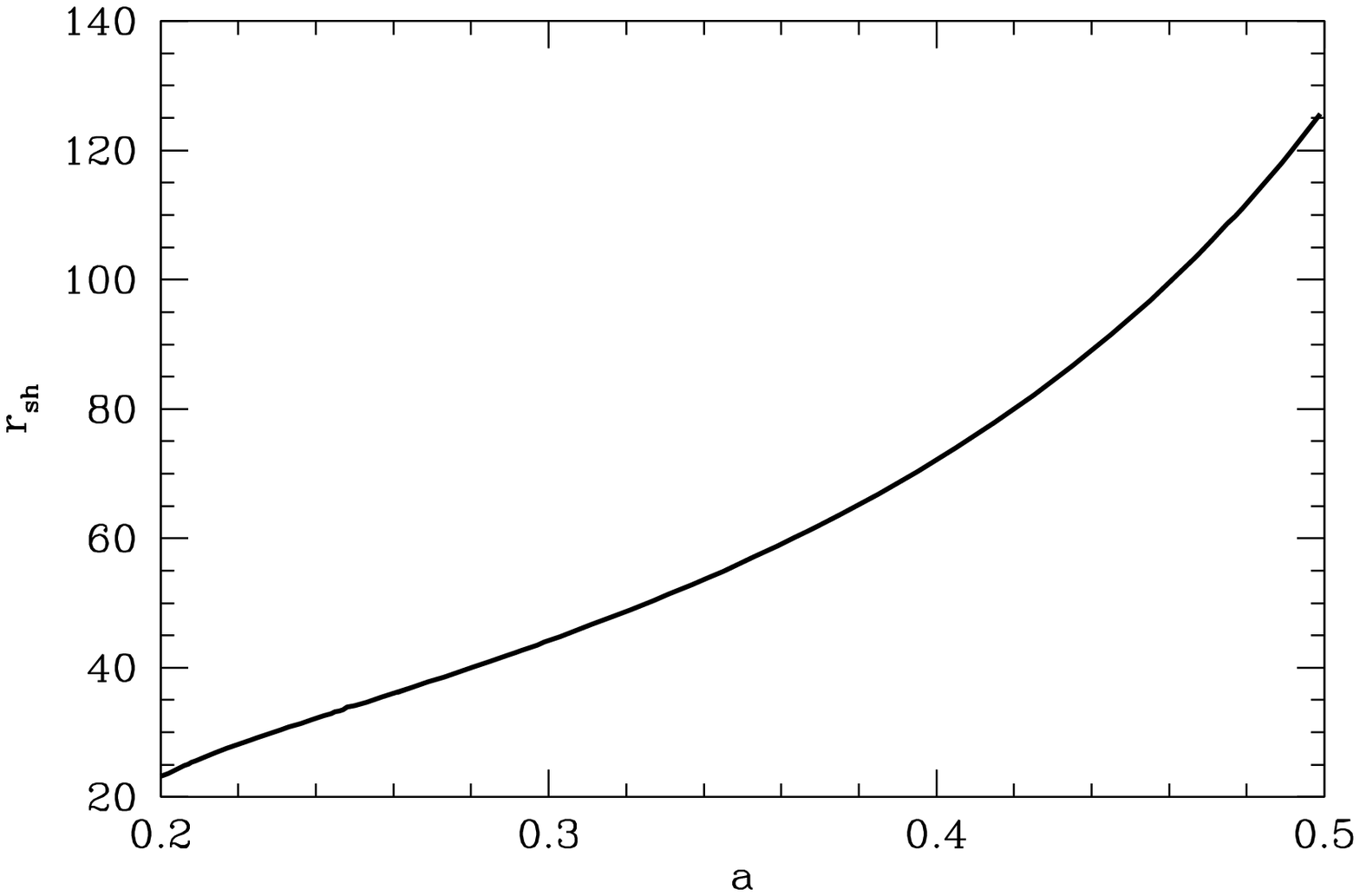,height=20cm,width=17.2cm,angle=0.0}}
\vskip -1.0cm
{\bf Figure 5a} 
}
\end{figure}
\begin{figure}
\vbox{
\vskip -8.0cm
\centerline{
\psfig{file=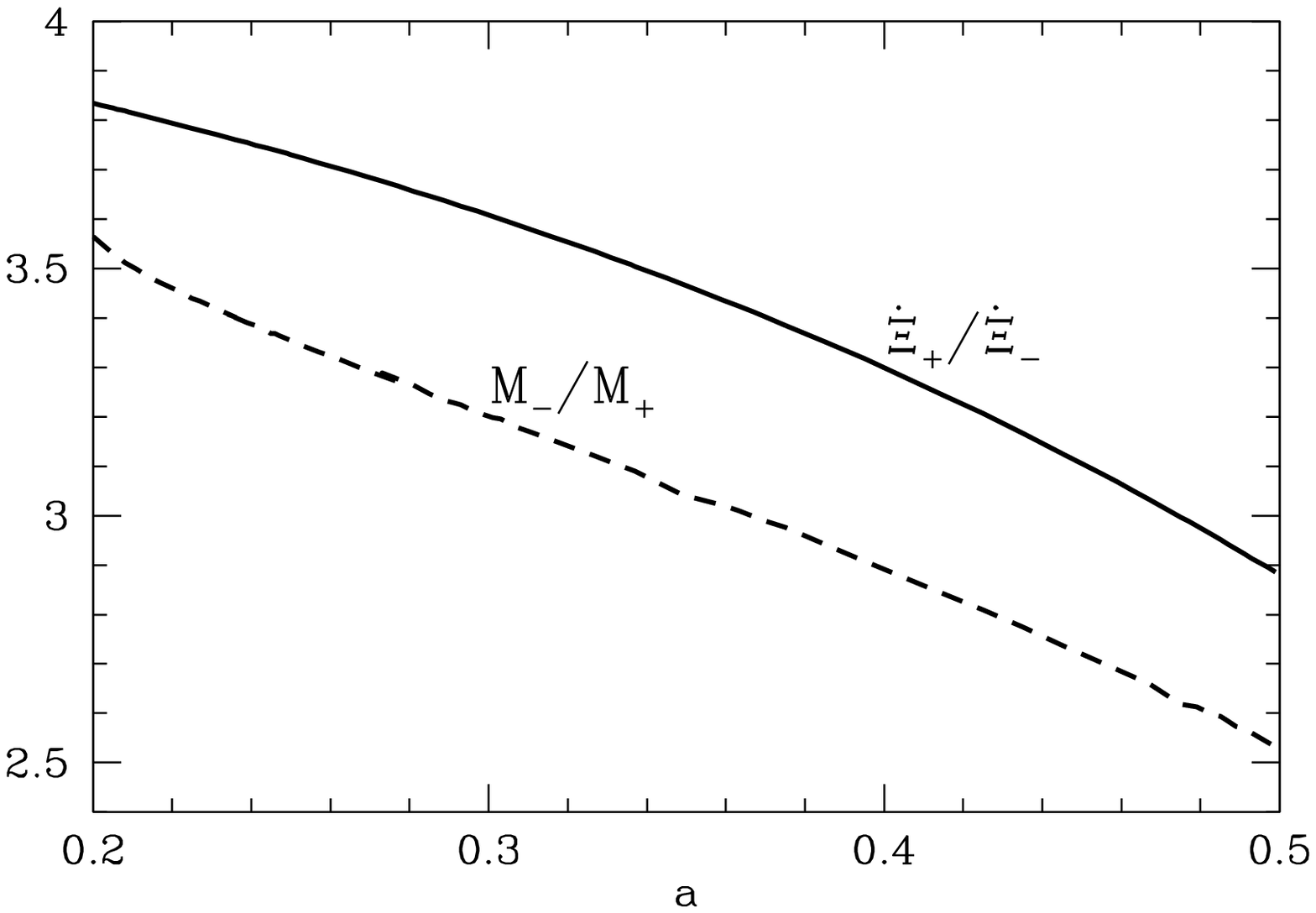,height=20cm,width=17.2cm,angle=0.0}}
\vskip -1.0cm
{\bf Figure 5b} 
}
\end{figure}
\begin{figure}
\vbox{
\vskip -9.0cm
\centerline{
\psfig{file=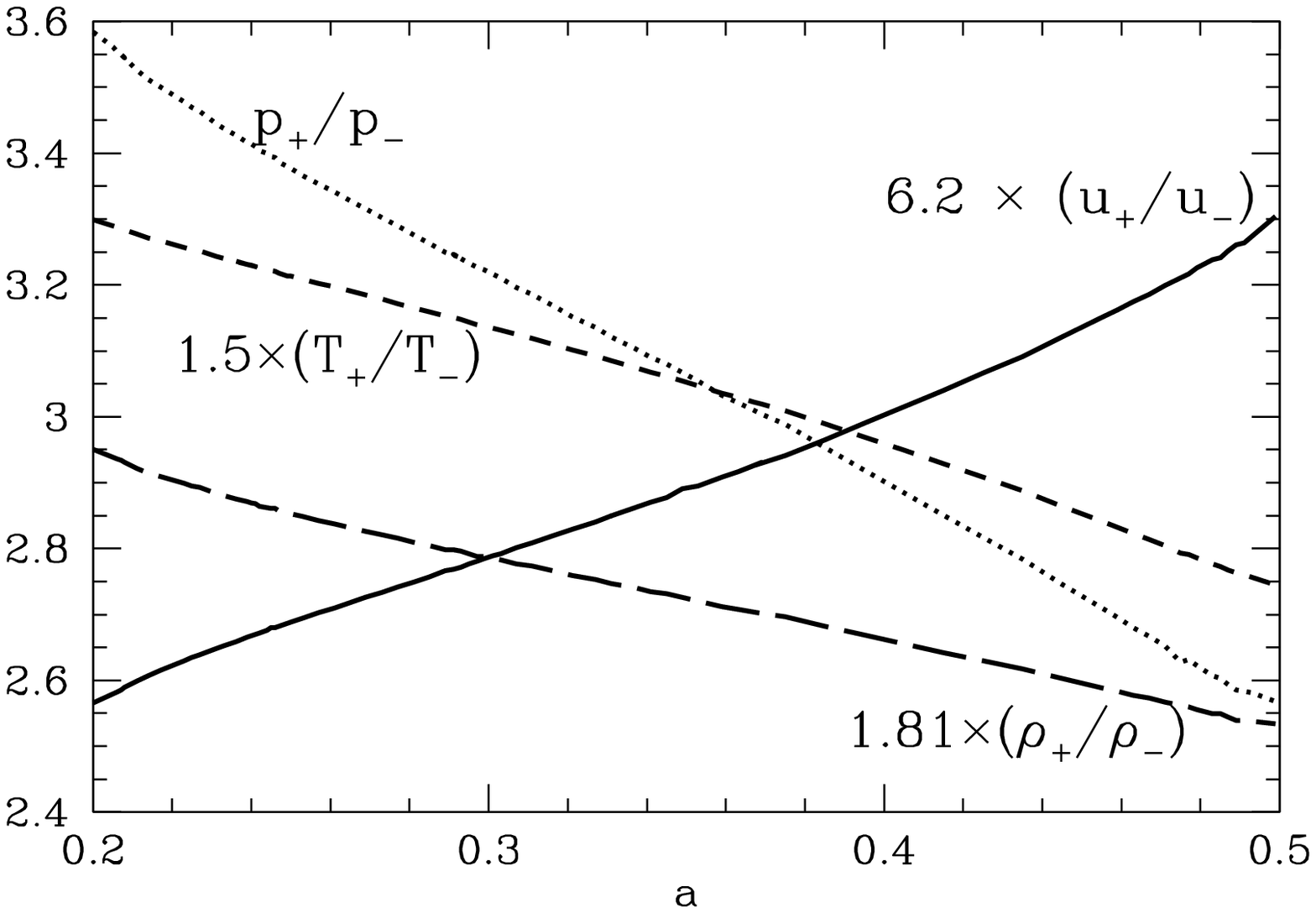,height=20cm,width=17.2cm,angle=0.0}}
\vskip -1.0cm
{\bf Figure 5c} 
}
\end{figure}
\begin{figure}
\vbox{
\vskip -9.0cm
\centerline{
\psfig{file=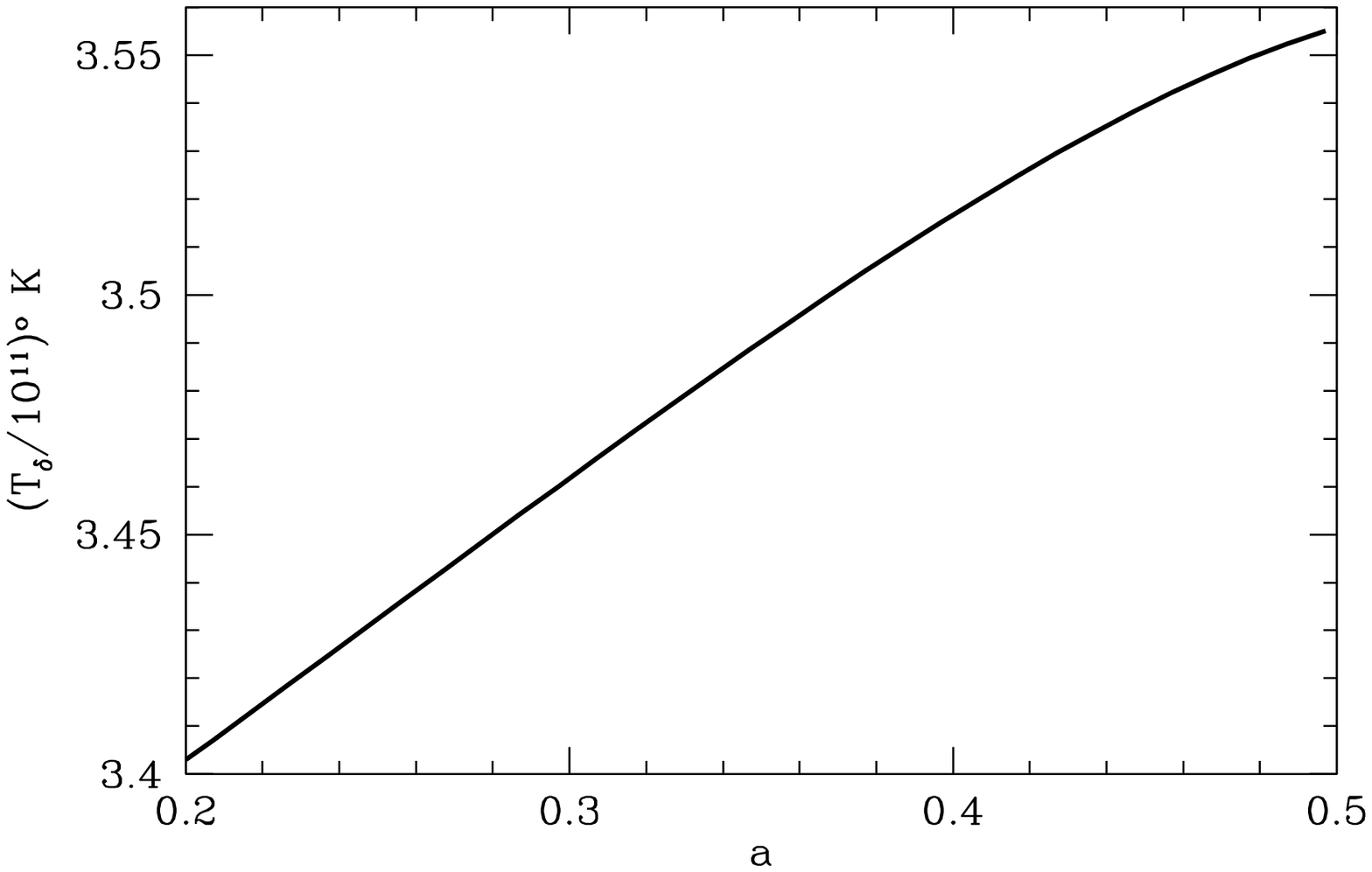,height=20cm,width=17.2cm,angle=0.0}}
\vskip -1.0cm
{\bf Figure 6a} 
}
\end{figure}
\begin{figure}
\vbox{
\vskip -8.0cm
\centerline{
\psfig{file=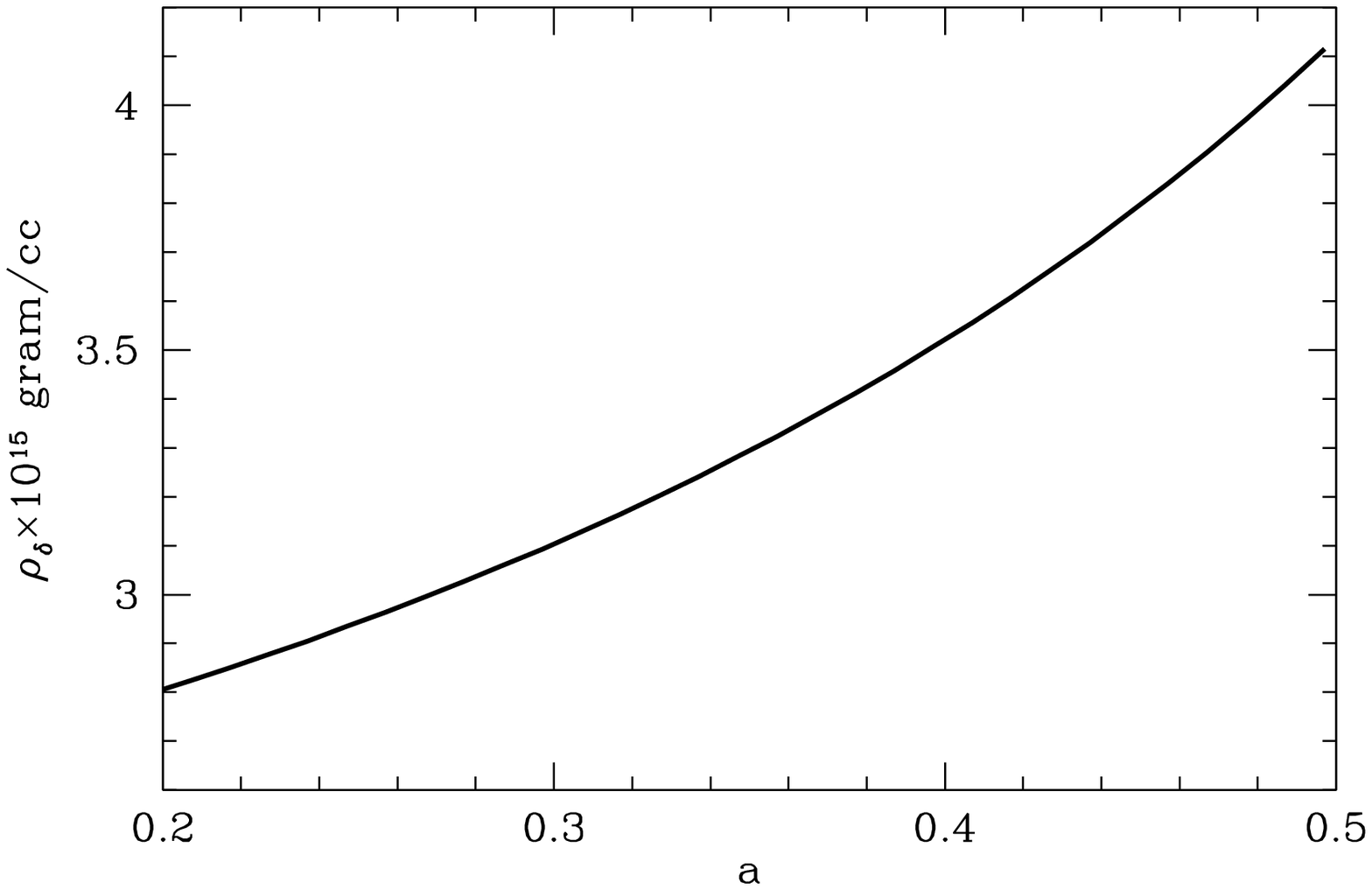,height=20cm,width=17.2cm,angle=0.0}}
\vskip -1.0cm
{\bf Figure 6b} 
}
\end{figure}
\begin{figure}
\vbox{
\vskip -8.0cm
\centerline{
\psfig{file=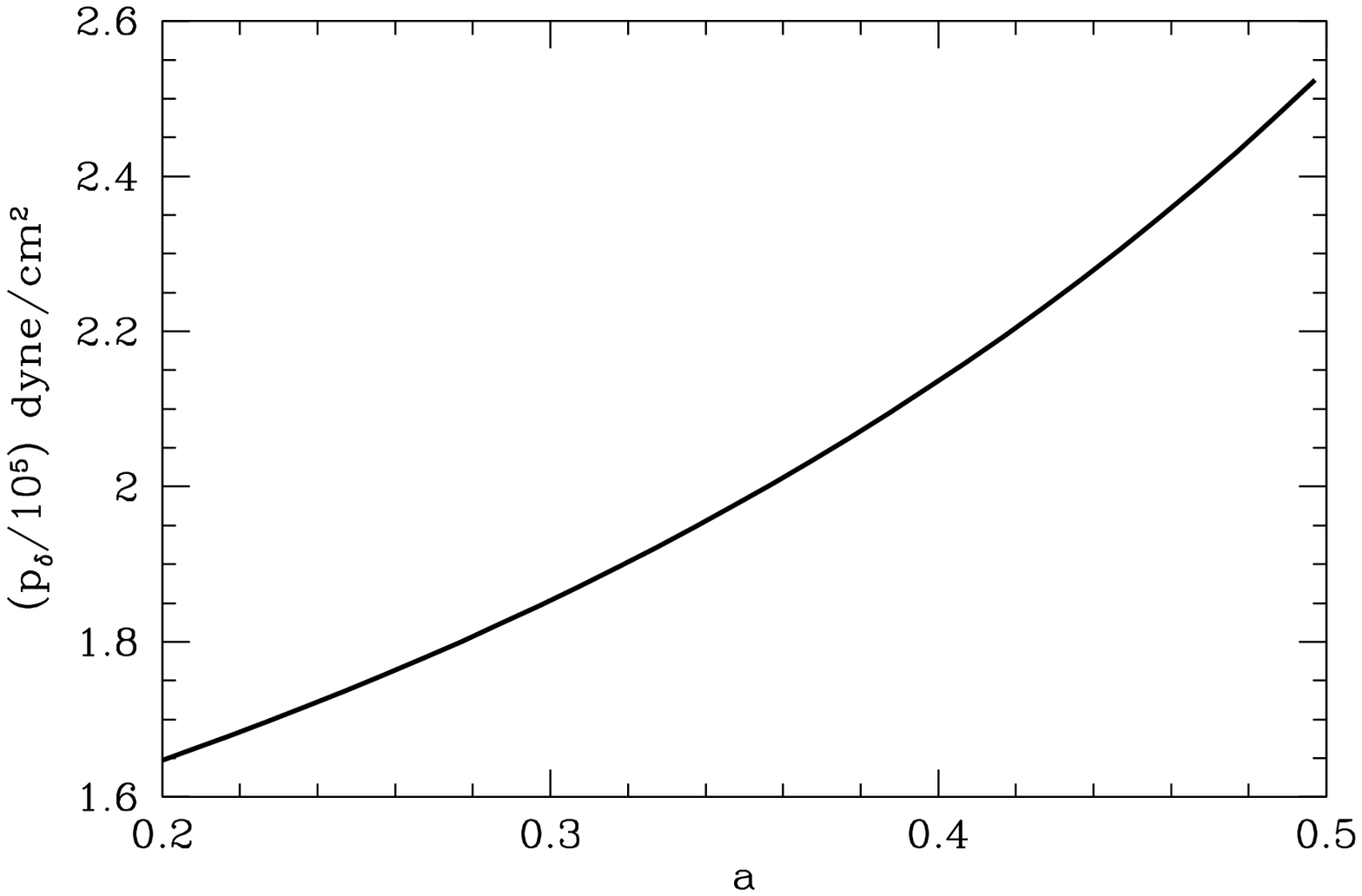,height=20cm,width=17.2cm,angle=0.0}}
\vskip -1.0cm
{\bf Figure 6c} 
}
\end{figure}

\end{document}